\documentclass[sigconf]{acmart}

\AtBeginDocument{%
  \providecommand\BibTeX{{%
    \normalfont B\kern-0.5em{\scshape i\kern-0.25em b}\kern-0.8em\TeX}}}

\copyrightyear{2024} 
\acmYear{2024} 
\setcopyright{rightsretained} 
\acmConference[CUI '24]{ACM Conversational User Interfaces 2024}{July 8--10, 2024}{Luxembourg, Luxembourg}
\acmBooktitle{ACM Conversational User Interfaces 2024 (CUI '24), July 8--10, 2024, Luxembourg, Luxembourg}\acmDOI{10.1145/3640794.3665887}
\acmISBN{979-8-4007-0511-3/24/07}

\usepackage{csquotes}

\begin{document}

    \title[A Provocation on the Utilisation of Persona in LLM-based Conversational Agents]{Building Better AI Agents: A Provocation on the Utilisation of Persona in LLM-based Conversational Agents}

\author{Guangzhi Sun}
\authornote{Both authors contributed equally to this research.}
\orcid{0000-0002-5886-056X}
\affiliation{%
  \institution{University of Cambridge}
  \city{Cambridge}
  \country{United Kingdom}
}  \email{gs534@cam.ac.uk}

\author{Xiao Zhan}
\authornotemark[1]
\orcid{0000-0003-1755-0976}
\affiliation{%
  \institution{King's College London}
  \city{London}
  \country{United Kingdom}}
\email{xiao.zhan@kcl.ac.uk}

\author{Jose Such}
\orcid{0000-0002-6041-178X}
\affiliation{%
 \institution{King's College London}
 \city{London}
 \country{United Kingdom}\\
 \institution{\& VRAIN, Universitat Politecnica de Valencia, Spain}
 }
\email{jose.such@kcl.ac.uk}

\renewcommand{\shortauthors}{Sun and Zhan et al.}

\begin{abstract}

The incorporation of Large Language Models (LLMs) such as the GPT series into diverse sectors including healthcare, education, and finance marks a significant evolution in the field of artificial intelligence (AI). The increasing demand for personalised applications motivated the design of conversational agents (CAs) to possess distinct personas. 
This paper commences by examining the rationale and implications of imbuing CAs with unique personas, smoothly transitioning into a broader discussion of the personalisation and anthropomorphism of CAs based on LLMs in the LLM era.

We delve into the specific applications where the implementation of a persona is not just beneficial but critical for LLM-based CAs. The paper underscores the necessity of a nuanced approach to persona integration, highlighting the potential challenges and ethical dilemmas that may arise. Attention is directed towards the importance of maintaining persona consistency, establishing robust evaluation mechanisms, and ensuring that the persona attributes are effectively complemented by domain-specific knowledge.


\end{abstract}

\ccsdesc[500]{Security and privacy~Social aspects of security and privacy}
\ccsdesc[500]{Security and privacy~Usability in security and privacy}
\ccsdesc[300]{Computing methodologies~Discourse, dialogue and pragmatics}
\ccsdesc[300]{Human-centered computing~HCI theory, concepts and models}
\ccsdesc[500]{Computing methodologies~Natural language processing}
\keywords{Large language model, persona, personality, conversational agent, ChatGPT, natural language processing}



\maketitle

\section{What Does `Persona' Mean in the Context of Conversational Agents?}



In the context of conversational agents (CAs), the concept of \emph{persona} represents the essence or `soul' of these agents. Persona encapsulates the distinct tone, voice, and personality that characterizes a CA, transforming mechanical interactions into engaging, human-like conversations~\cite{sutcliffe2023survey,kim2019designing}. Commonly, these attributes of persona can consist any type of information that intend to capture personal characteristics about an individual~\cite{liu2022persona}, and are relatively static (race), and slowly change over time (age), or temporary (emotional status)~\cite{li2016persona,yang2019computational}.


Before delving deeper into the discussion of personas in CAs, it's important to distinguish this concept from the idea of `personality' that has been explored in prior research~\cite{lessio2020toward,liao2020racial,pradhan2021hey,Alexapersonality}. While personality traits, such as being "friendly" or "smart," or frameworks like the Myers-Briggs Type Indicator (MBTI)~\cite{Briggs1987}, might define certain characteristics shared by groups of individuals, a persona in CAs represents a more complex and consistent identity~\cite{pradhan2021hey,zhang2018personalizing}. This persona transcends mere personality traits, serving as an external manifestation of a character's unique identity. For instance, when a CA is designed with the persona of a specific character, say, Sherlock Holmes, it consistently embodies the unique attributes and behaviors of that character throughout interactions. This specificity differs significantly from assigning generic traits like 'bravery' and 'smartness' to a CA. In the latter case, the CA might alternate between different characters who share these traits, such as both Sherlock Holmes and Hermione Granger, depending on the context of the interaction. Thus, the persona of a CA is a more nuanced and stable layer that defines its interaction style and character representation.

\subsection{Persona in CAs in pre-LLM era}
Recent research in the field of CAs has focused extensively on enhancing the capabilities of chatbots, aiming to imbue them with more human-like characteristics. This initiative is driven by the goal to significantly boost user engagement, among other benefits. The development of a persona for CAs such as chatbots has emerged as a key strategy in this domain. The introduction of these personas is a testament to the evolving sophistication of chatbot technology, reflecting a deeper understanding of human-chatbot interaction dynamics~\cite{hwang2021applying}. 

Two main avenues of exploration have emerged: the technical research stream pushes the boundaries of what is technically possible~\cite{zhou2020design,danielescu2018bot,liao2020racial,li2016persona,sordoni2015neural,vinyals2015neural,sutcliffe2023survey}, the social research stream ensures that these advancements are grounded in a thorough understanding of user needs, preferences, and the broader societal context~\cite{rashkin2018towards,zhong2020towards,bickmore2010response,hwang2021applying}. 

\subsubsection{Technical research.} 
Previous studies have proposed various methods for embedding personas into traditional chatbots\footnote{Unlike our approach that distinctly separates persona from personality, some prior research conflates these concepts without addressing their nuances. Therefore, the summary in this section includes works that focus on `personality' as well.}. The categories used are broad --- for a comprehensive summary of the model and a survey see \cite{sutcliffe2023survey}. The more widely known examples are that 
neural models of conversation generation provided a simple mechanism for incorporating personas as embeddings \cite{li2016persona,sordoni2015neural,vinyals2015neural}. More recently, \citeauthor{liao2020racial} created personas for conversational agents that had distinct gender and race to understand user preferences~\cite{liao2020racial}. As one example of a project that is guided by user data, persona XiaoIce was designed based on a large scale analysis of human conversations~\cite{zhou2020design}. In doing so, the designers found that the majority of “desired” users are young and female. Hence, they designed XiaoIce’s persona around an “18-year-old girl”~\cite{zhou2020design}. As another example, Danielescu and Christian~\cite{danielescu2018bot} designed personas for a conversational coaching system where they involved customers by interviewing them and brainstorming with them, finding that their preferences may vary based on their culture and region.


\subsubsection{Social research.} The academic community has consistently maintained a positive attitude towards endowing chatbots with personas. Incorporating a distinct persona in CAs significantly influences the development of a robust relationship in human-agent interactions. It has been demonstrated that a well-crafted persona can significantly enhance the capacity of CAs to engage in empathetic conversations~\cite{rashkin2018towards,zhong2020towards}. This is mirrored from empirical research, such as that by \citeauthor{zhong2020towards}~\cite{zhong2020towards}, has established the role of persona in fostering empathy within human conversations from the psychological perspective. Moreover, the positive contribution of persona is recognised in specific areas such as healthcare where CAs assume varied roles. For instance, \citeauthor{bickmore2010response}~\cite{bickmore2010response} found that an empathetic persona in an agent is effective for managing mental health, whereas an agent with a subtle persona guiding exercise can enhance commitment to behavior change. Similarly, preliminary research conducted in \cite{hwang2021applying} indicates that chatbots embodying roles like doctors, in comparison to generic bots, achieve higher user acceptance, intimacy, and trust in healthcare-related interactions.



\section{Reality or Aspiration?}



\begin{figure}
    \centering
    \includegraphics[width=0.46\textwidth]{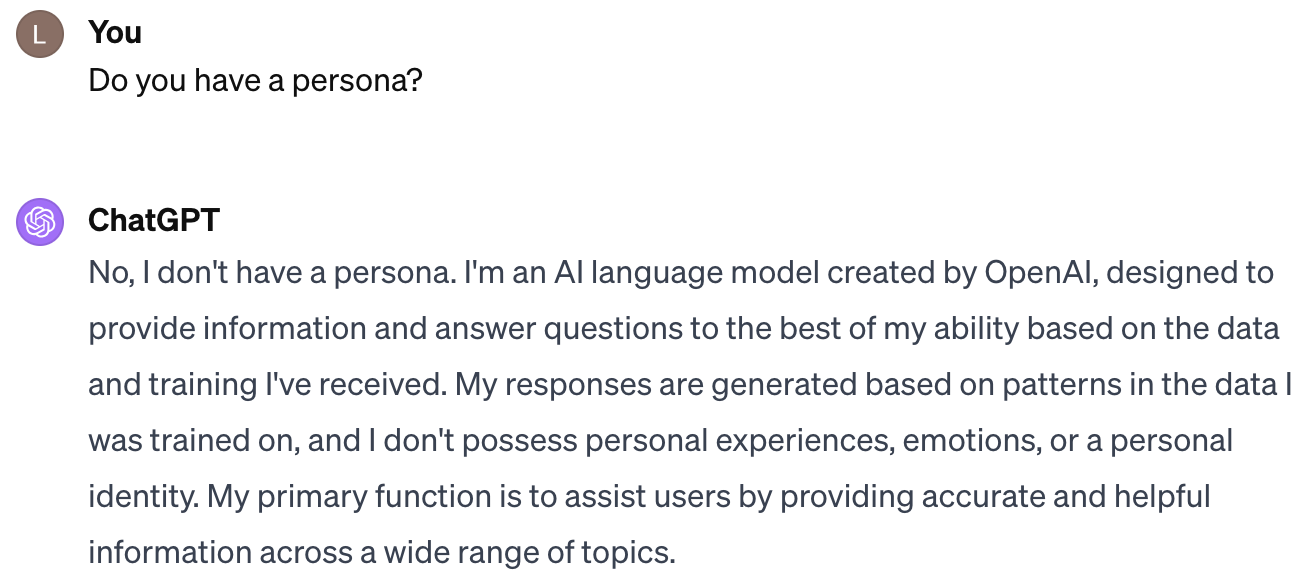}
    \caption{A screenshot of a dialogue with GPT-4-0125-preview. This suggests that GPT-4 does not embody a specific persona. However, this conclusion is based on the model's output, which may not fully align with the designers' intentions.} 
    \label{fig:persona-chatgpt}
\end{figure}

Large Language Model (LLM)-based CAs, exemplified by systems like ChatGPT\footnote{\url{https://openai.com/chatgpt}}, are rapidly being integrated into various critical sectors, underscoring their growing significance in practical applications. These include, but are not limited to, healthcare~\cite{cascella2023evaluating,lai2023supporting,thirunavukarasu2023large}, education~\cite{xiao2023exploratory,kohnke2023chatgpt,mbakwe2023chatgpt}, and finance~\cite{lakkaraju2023can,lakkaraju2023llms,wu2023bloomberggpt}, among others.

These LLM-based CAs, which are originally developed for general-purpose applications, do not prioritize the establishment of a distinct persona during their design phase. For example, as illustrated in Figure~\ref{fig:persona-chatgpt}, ChatGPT, a typical instance of such systems, is structured to function without a predefined persona, focusing instead on delivering information and interaction capabilities that are broadly applicable across various contexts and user requirements\footnote{Despite this observation, no official documentation or evidence has been found to indicate that ChatGPT was deliberately designed to incorporate distinct personas.}.

Nevertheless, the integration of personas in LLM-based CAs should not be viewed as an unattainable goal. Online resources (including blogs~\cite{chatgptpersona,createpersona}, and technical reports~\cite{white2023prompt}) already provide guidance on designing specific personas to optimize ChatGPT's effectiveness across various roles, typically achieved by customizing initial conversation prompts to assign a desired persona. Concurrently, numerous empirical studies~\cite{jiang2023personallm,durmus2023towards,kong2023better,zhou2022large,chan2023chateval,park2023generative,park2022social,argyle2023out} have examined and demonstrated the practicality of assigning personas to LLM-based CAs. Among them, some promising results indicated that endowing LLM-based CAs with personas leads to satisfactory outcomes. These include the ability to express opinions similar to people from some controes~\cite{durmus2023towards}, offering useful answers~\cite{kong2023better}, team working~\cite{chan2023chateval}, and enhancing the overall truthfulness in their responses~\cite{zhou2022large}.

However, upon deeper analysis of persona-based CAs, it becomes evident that LLM-based CAs are still far from embodying specific personas at this stage, highlighting a substantial developmental path that lies ahead. For instance, significant performance disparities exist between different GPT versions. Stories from GPT-4 personas are generally more readable, coherent, and believable, while ChatGPT tends to deviate from the provided prompts, failing to adhere strictly to the prescribed personas~\cite{jiang2023personallm}. In \cite{shu2023you},  a study was conducted to assess if the prevailing prompt-based approach facilitates LLM-based CAs in delivering consistent and robust responses. Their investigation, which included testing 15 open-source LLMs, ultimately revealed that most models lacked a consistent persona. Furthermore, it's noteworthy that malicious actors sometimes exploit these characteristic, manipulating them to generate toxic responses\cite{deshpande2023toxicity,zhuo2023exploring}.

\section{Persona needs in LLM-based CAs}


In the current landscape dominated by LLMs, the importance of persona has not diminished, rather, it often takes on an even more critical role. In this section, we will explore various situations and use cases where the persona of a LLM-based CA is particularly crucial.

\subsection{Participant Simulation}


\citeauthor{hagendorff2022machine}~\cite{hagendorff2022machine} conducted an evaluation of GPT-3.5 through cognitive response tests and discovered that the error patterns of the language model qualitatively reflect intuitive behaviors akin to those found in humans. Furthermore, it often fails in similar reasoning tasks as humans do~\cite{dasgupta2022language}. These findings underscore the significant potential of LLMs in capturing aspects of human behavior. Based on these findings, LLMs are increasingly being considered and used to simulator human beings with different personas. Recent studies~\cite{argyle2023out,aher2023using,park2022social} have provided substantial evidence that LLMs simulating user responses can replicate social science experiments and online forums with a high degree of consistency comparable to those obtained using actual human participants.

The future of simulating various user types appears brighter as the accuracy of such simulations continues to improve. Experiments and studies in fields constrained by traditional methodologies stand to benefit significantly from advanced technologies like LLMs. For example, research exploring interactions with individuals who have mental health issues often faces ethical dilemmas and heightened risk assessments. Utilizing LLMs equipped with well-defined personas to simulate such participants can expedite research processes while minimizing potential risks to the interaction between researchers and subjects. Additionally, in studies seeking diverse and balanced samples, recruitment challenges often arise, especially when targeting specific demographic backgrounds. LLMs can be programmed to represent a range of demographics and personas, thus addressing recruitment limitations efficiently. Moreover, the financial implications of user studies involving large participant groups are considerable. By incorporating personas into LLMs, researchers can conduct extensive studies more cost-effectively, without compromising the breadth and diversity of participant profiles.

\subsection{Role Playing in Specific Domains}



LLM-based CAs, when programmed with specific personas, offer substantial support to educators, especially teachers, in improving their development of educational content, enriching their teaching methodologies, and bolstering their self-assurance. For instance, such agents can simulate a variety of student personas, enabling teaching assistants (TAs) to engage in realistic interaction scenarios~\cite{markel2023gpteach}. This approach allows TAs to refine their skills in providing feedback and effectively addressing the needs of students with diverse characteristics, learning goals, and educational backgrounds. This comprehensive and authentic practice environment is instrumental in equipping TAs with the necessary competencies to minimize instructional mishaps in real-world teaching situations. Similarly, they have the potential to significantly enhance the professional skills of lawyers, physicians, and other specialists. These LLM-based CAs can simulate interactions with diverse patient types, including the elderly and those with unique symptoms or needs. Traditionally, such simulations form a crucial part of training before professionals are fully qualified. Now, with the integration of LLM agents equipped with specialized personas, this training phase can be streamlined and made more intelligence-oriented, offering a sophisticated approach to professional skill development.

Beyond their assistive role, these technologies can have personas to simulate domain experts, notably in healthcare, education and law. Here, CAs would blend intellectual and emotional support, innovatively simulating roles such as caregivers, tutors, and legal advisors. 
Nonetheless, their effectiveness hinges also on having accurate, domain-specific expertise, a critical aspect we will discuss in Section~\ref{sec:more-than-persona}.

\subsection{Brand Representation}

The persona of an LLM-based CA plays a crucial role in brand representation by aligning with the brand's values, enhancing user engagement, and serving as a differentiator in a crowded market. For instance,

\begin{displayquote}
`` Domino's pizza created `Dom', a virtual ordering assistant. Dom's persona is friendly and efficient, reflecting the brand's focus on convenient and fast service. Dom allows customers to order pizza using conversational language, making the process more engaging and aligning with Domino's commitment to innovation in delivery and customer service.''~\cite{Dom} 
\end{displayquote}

A well-defined persona ensures that the agent's communication style and tone are consistent with the brand's identity, fostering a stronger and more coherent brand image. This alignment is essential not only for maintaining brand consistency but also for creating a more engaging and relatable experience for users. In an environment where many companies employ similar technologies, a distinctive persona can significantly set a brand apart, making it more memorable and appealing to customers. This unique identity helps in building customer loyalty and establishing a competitive edge.

\section{Challenges and Caveats}

\subsection{Consistency Is the Top Priority}

The primary objective in the design of CAs is to establish and nurture a robust connection with users, facilitating ongoing engagement over extended periods~\cite{shum2018eliza}. Achieving this necessitates the ability of the CAs to engage in sustained, meaningful conversations~\cite{yan2016attribute2image,song2019exploiting}. Recent findings~\cite{garcia2023not,sclar2023quantifying} indicated that LLM-based CAs exhibit a heightened sensitivity to subtle and sensitive words within the context, leading to inconsistent outputs. This characteristic has raised concerns about the ability of LLM-based CAs to maintain a \textbf{consistent persona}\footnote{Consistency and coherency: \textbf{Consistency} means whether elements of persona remain unchanged throughout the conversation, e.g. you can not be a kid in one turn while talking like an old person in another. \textbf{Coherency} refers more to whether the persona elements are coherent, e.g. you can not say something like "I went on a trip with my wife for my 5-year-old birthday". Consistency cares more about persona across different turns, i.e. evolution across time and can only be defined for multi-turn dialogue.} throughout multiple dialogue exchanges~\cite{li2015diversity,jiang2023personallm}. This observation underscores the challenge of ensuring that these AI systems not only understand and process language effectively but also retain a consistent and contextually appropriate persona over successive interactions. Moreover, as discovered in~\cite{vinyals2015neural}, having inconsistency of persona is one of the major obstacles in achieving the long-term objective of developing human-like CAs to pass the Turing test~\cite{turing2009computing} 
Addressing this issue is critical in enhancing the reliability and user trust in conversational AI technologies~\cite{lessio2020toward,moussawi2021effect}.

Unfortunately, most widely-used LLMs struggle to align responses consistently with latent persona attributes~\cite{shu2023you}. This inconsistency is particularly evident in complex tests, like reversing question meanings using negation. Only two out of fifteen models tested in this paper, 
achieved some level of consistency~\cite{shu2023you}, 
highlighting the need for further development to enhance persona consistency in LLM responses.




\subsection{Are There Effective Ways to Evaluate Persona and Its Consistency?}


So far, a systematic approach to evaluate and verify persona application in LLM-based CAs has not been established. However, there exists some noteworthy attempts, such as employing empirical frameworks for indirectly assessing the persona of CAs~\cite{safdari2023personality,shu2023you,hagendorff2023machine}. This can be achieved through psychometric testing or by analyzing survey results. 

It appears that one cannot ascertain the specific persona a LLM-based CA is exhibiting simply by prompting queries such as "what is your persona" or "describe your persona." Consider a scenario where an LLM-based CA is programmed or processed to embody a certain persona. The reality is, people cannot exhaustively enumerate all the traits of this persona, leaving room for the CA to exhibit some degree of self-expression in its responses. Moreover, the inherent unpredictability of the LLM adds a layer of complexity. For instance, the CA might be defined as "a 21-year-old physics student from Canada with a particular temperament..." but these specifications are insufficient to confine it to a specific character or individual. For example, in one interaction round, the LLM-based CA may fit this description but have a preference for bowling, while in the next round, it might have the same foundational characteristics but prefer skiing. In this situation, its persona has changed, yet such changes are subtle and challenging to detect and define. We can only ascertain their adherence to our initial constraints through certain predetermined questions. The CA might perfectly execute the task, but when asked about other aspects, like hobbies, it might reveal inconsistencies. Such situations are unpredictable and difficult to capture.

Moreover, we acknowledge that individual perceptions of a system's persona can vary. For instance, a chatbot with a female-like voice might be considered by someone as sufficiently demonstrating a persona. This variability poses challenges in establishing a universally accepted standard for assessing a system's capability to exhibit a persona.




\subsection{More than Persona}
\label{sec:more-than-persona}

This point becomes particularly prominent in our discussion about endowing "characters" with the ability to play different domain experts in LLM-based CAs. We believe that for CAs to successfully assume the required roles, it is essential to impart not only fundamental character traits such as demographics, age, and gender but also corresponding knowledge. Characters must also be equipped with professional knowledge that aligns with their identities. For example, a character role as an ophthalmologist should be familiar with basic ophthalmology as well as be able to fluently address complex questions about eye diseases. Similarly, a character claiming to be a judge should be acquainted with basic legal statutes. Moreover, effective role-playing entails not just possessing knowledge but also the ability to adapt responses according to different contexts. For instance, when asking a business consultant character about market trends, it should be capable of considering the current economic environment and specific industry dynamics to provide informed responses. This approach elevates that researchers should always think more than just persona. In designing and implementing these roles, careful consideration must be given to their expertise and adaptability to ensure their effectiveness and credibility in their respective fields.

Hallucination, as one of the crucial caveats in most LLMs, will trigger new problems in persona-based LLMs. Persona-based LLMs may hold the wrong belief in certain facts about themselves, e.g. occupation and social relationships. Current hallucination detection or evaluation methods depend on fixed non-persona-based datasets or uncertainty and inconsistency measures\cite{manakul2023selfcheckgpt,sun2024crosscheckgpt}. However, in persona-based LLMs, such beliefs, once established, tend to remain consistent where the model is confident. Therefore, high self-consistency in persona-based LLMs requires more customised hallucination detection and prevention approaches to be developed.

\subsection{Ethical Considerations}

As with any AI-based technology, integrating personas into LLM-based CAs presents a dual-edged sword. It offers significant benefits but can also harbor potential risks. The use of such technology inevitably necessitates careful consideration of ethical issues. Potential harms include, but are not limited to, the ethics of deception and the reinforcement of societal stereotypes. We encourage our audience to refer to the provocation paper~\cite{pradhan2021hey} for a more comprehensive discussion on the ethical considerations surrounding the use of personas in these systems. Here we refrain from redundant elaboration of previously stated points.


\section{Conclusion}



In integrating persona into LLM-based CAs, this provocation highlights the significance of persona to enhance human-like interactions. It covers the criticality of various applications, and meanwhile puts forward challenges in achieving persona consistency and domain-specific adaptability. In conclusion, although the prospect of creating CAs with high effectiveness and human resemblance is promising, prioritizing ethical standards and tackling technical challenges is essential. Future efforts must aim for responsible development that maximizes the benefits of persona integration while addressing its complexities.



    

\begin{acks}
We thank CUI’s anonymous reviewers for their constructive comments on previous drafts of this paper. This research was partially funded by EPSRC under grant \emph{SAIS: Secure AI assistantS} (EP/T026723/1) and by the INCIBE's strategic SPRINT (Seguridad y Privacidad en Sistemas con Inteligencia Artificial) C063/23 project with funds from the EU-NextGenerationEU through the Spanish government's Plan de Recuperación, Transformación y Resiliencia.
\end{acks}

\bibliographystyle{ACM-Reference-Format}
\bibliography{sample-base}


\appendix

\end{document}